\documentclass[12pt]{article}

\usepackage{amssymb,amsmath}
\usepackage{epsfig}

\begin{document}
\def\lf{16\pi^2}
\def\Tr{{\rm Tr}}

\pagenumbering{arabic}

\begin{titlepage}

\begin{flushright}
hep-th/0204045 \\
WIS/14/02-APR-DPP
\end{flushright}

\vspace{7 mm}

\begin{center}
 {\huge Exactly Marginal Deformations}
 \vspace{3 mm} {\huge of ${\cal N}=4$ SYM and}
\vspace{3 mm} {\huge of its Supersymmetric Orbifold Descendants}
\end{center}
\vspace{10 mm}
\begin{center}
{\large

Ofer Aharony\footnote{{\tt Ofer.Aharony@weizmann.ac.il}. Incumbent of
the Joseph and Celia Reskin career development chair.} and
Shlomo S.~Razamat\footnote{{\tt Razamat@wisemail.weizmann.ac.il}.}
}\\
 \vspace{3mm} Department of
Particle Physics,\\ The Weizmann Institute of Science,\\ Rehovot
76100, Israel\\

\end{center}
\vspace{7mm}
\begin{center}
{\large Abstract}
\end{center}
\noindent
 In this paper we study \textit{exactly} marginal deformations of
 field theories living on D3-branes at low energies. These theories include
  ${\cal N}=4$ supersymmetric Yang-Mills theory
 and theories obtained from it via the orbifolding procedure.
   We restrict ourselves only to orbifolds and deformations
 which leave some supersymmetry unbroken. A number of new families of
 ${\cal N}=1$ superconformal field theories are
 found. We analyze the deformations perturbatively, and also by using
 general arguments for the dimension of the space of \textit{exactly} marginal
 deformations. We find some cases where the space of perturbative
 \textit{exactly} marginal deformations is smaller than the prediction of the
 general analysis (at least up to three-loop order), and other cases where
 the perturbative result (at low orders) has a non-generic form.

\vspace{7mm}
\begin{flushleft}
April 2002
\end{flushleft}
\end{titlepage}

\section{Introduction}

In recent years there has been a great burst of research and interest
in ${\cal N}=4$ Super Yang-Mills (SYM) theory. Much of
the interest is due to the fact that this theory has a string theory
dual via the AdS/CFT correspondence (see
\cite{a:Maldac,a:OA}). ${\cal N}=4$ $SU(N)$ SYM appears in this
context as a low-energy effective description of $N$ coincident D3-branes.
By looking at D3-branes at an orbifold point one can obtain
\cite{a:quiver,a:quiver2,a:SilKach}
effective descriptions in terms of conformal field theories with less
supersymmetries (${\cal N}=2$, ${\cal N}=1$, and perhaps also
${\cal N}=0$). It is interesting to look
at \textit{exactly} marginal deformations of such theories. In the
AdS/CFT \cite{a:Maldac} correspondence such deformations on the field
theory side correspond
to moduli of the string theory. For instance \cite{a:SilKach},
in the ${\cal N}=2$ $\mathbb{Z}_k$
orbifold theory there are $k$ \textit{exactly} marginal deformations which
preserve the ${\cal N}=2$ SUSY. These deformations correspond to the
string coupling and ALE blow-up modes on the string theory side.

    In this work we investigate marginal deformations, which preserve at
    least ${\cal N}=1$ supersymmetry, of ${\cal N}=4$ SYM and its
    orbifold ${\cal N}=2$ and ${\cal N}=1$ descendants. There are two known
    \textit{exactly} marginal deformations of this type for ${\cal
    N}=4$ SYM (see \cite{a:Leigh} and references therein)\footnote{Marginal
deformations of ${\cal N}=4$ SYM were also discussed in \cite{a:Leigh1,a:Leigh2}},
and we show that these (and the gauge coupling)
are the only \textit{exactly} marginal supersymmetric
deformations of this theory.
    The planar diagram contribution in the orbifold theories is the same as
    in the ${\cal N}=4$ theory \cite{a:N_plan1,a:N_plan2}, so in the
    large $N$ limit many correlation functions in these theories
    coincide up to some gauge coupling rescaling. Thus, one could expect
    that the orbifold theories possess similar \textit{exactly}
    marginal deformations. We will show that this is actually the case
    even without going to the large $N$ limit,
    and we find additional \textit{exactly} marginal operators from the
    twisted sectors. In some cases we find that the dimension of the
    space of \textit{exactly} marginal deformations at low orders in
    perturbation theory is smaller than the general analysis implies.
    For $SU(N=3)$ gauge groups more deformations are possible, and
we find a much larger
    number of \textit{exactly} marginal deformations.

One motivation for studying such families of conformal theories is
that they can then be used as a starting point for renormalization group
(RG) flows by turning
on additional relevant operators. For instance, one might hope that by
starting from a theory obtained by an \textit{exactly} marginal deformation of
${\cal N}=2$ orbifold theories, one could flow to duality cascades of
the type considered in \cite{a:KleStr}\footnote{O.A. would like to
thank I. Klebanov and J. Polchinski for discussions on this
issue.}, for which a direct field theory definition is not known.
Unfortunately, we do not find any deformations which are useful for this.

The reason that we constrain ourselves to working with supersymmetric
field theories is the existence of relations between
$\beta$-functions in these theories, due to non-renormalization theorems. The
$\beta$-functions in supersymmetric field theories can be
expressed in terms of the anomalous dimensions, $\gamma$. For the
superpotential couplings this is a consequence of the superpotential
non-renormalization theorem, and for the gauge coupling it is the
NSVZ $\beta$-function
\cite{a:NSVZ1,a:NSVZ2,a:NSVZ3}.  For a
superpotential $W={1\over6}Y^{ijk}\Phi_i\Phi_j\Phi_k$ the beta function is
given by
\begin{equation} \label{betaY}
\beta_{Y^{ijk}}= Y^{p(ij}\gamma^{k)}_p =
Y^{ijp}\gamma^k_p+(k\leftrightarrow i)+(k\leftrightarrow j),
\end{equation}
and the NSVZ $\beta$-function for the gauge coupling
is\footnote{Here $Q$ is the one loop
gauge $\beta$-function, $C(R)^i_j = (R_A R_A)^i_j$ where $R_A$ is
the representation of the matter chiral superfields, $r=\delta_{AA}$,
and $C_1\delta _{AB} = f_{ACD}f_{BCD}$. }:

\begin{equation}\label{NSVZ_b}
\beta_g = {{g^3}\over{\lf}}\left[ {{Q- 2r^{-1}\Tr\left[\gamma
C(R)\right]}
    \over{1- 2C_1g^2{(\lf)}^{-1}}}\right].
\end{equation}

The strategy of our search for \textit{exactly} marginal
deformations is the following. We first make a generic computation of the
expected dimension of the space of \textit{exactly} marginal
deformations by using the relations between the $\beta$-functions
and the anomalous dimensions above, as in \cite{a:Leigh} and references
therein. Then we check
in perturbation theory whether these deformations actually appear,
and whether their form agrees with the general analysis.

In the next section we study \textit{exactly} marginal
deformations of ${\cal N}=4$ SYM theory. In section ~\ref{sec_orb} we
describe the orbifold procedure and study the \textit{exactly}
marginal deformations of the orbifold theories. The ${\cal N}=2$
case is studied in detail in \S\ref{sec_N_2}. Details of other cases
may be found in \cite{a:Thesis}.

\section{${\cal N}=4$ Super-Yang-Mills Theory}

Four dimensional ${\cal N}=4$ $SU(N)$ SYM theory
appears as a low energy description of the
physics on $N$ coincident D3-branes in type IIB superstring theory.
In ${\cal N}=1$ superspace notation
the matter content is three chiral superfields, $\Phi^i$, in the adjoint
representation of the gauge group. Besides changing the gauge coupling,
the only classically marginal deformations (for $N > 2$)
preserving ${\cal N}=1$ SUSY are superpotentials of the form
\begin{eqnarray}
  \frac{i\lambda\sqrt{2}}{3!}
     \epsilon_{ijk}\Tr({\Phi}^i[{\Phi}^j,{\Phi}^k])\nonumber,\\
  \frac{h_{ijk}}{3!}\Tr({\Phi}^i\left\{{\Phi}^j,{\Phi}^k\right\}).
\end{eqnarray}
Traces are taken in the fundamental representation of the gauge
group. In the ${\cal N}=4$ theory the $\lambda$ coupling is equal to
the gauge coupling and all $h_{ijk}$ vanish.

 From (\ref{betaY}),(\ref{NSVZ_b}) we obtain:
\begin{eqnarray}
\beta_g,\beta_{\lambda}\propto \Tr(\gamma)\nonumber,\\
\beta_{h_{ijk}}\propto h_{p(ij}\gamma_{k)}^p.
\end{eqnarray}
If we turn on only $\lambda$,
$h_{111}=h_{222}=h_{333}$ and $h_{123}$, then
$\beta_{h_{ijk}}\propto Tr\gamma$ as well, and thus the single equation
$\Tr(\gamma)=0$ is enough to ensure conformal invariance.
We get one equation for
four coupling constants, so we expect a three dimensional manifold of
fixed points \cite{a:Leigh}.
We will see that these are the only \textit{exactly}
marginal supersymmetric deformations in this theory (up to the global $SU(3)$
symmetry we have between the three $\Phi^i$).

 The one-loop calculation of the beta functions and the anomalous
dimensions gives\footnote{Here $C_2\delta_{ab}\equiv \Tr(T_aT_b)$
where $T_a$ are $SU(N)$ generators in the fundamental representation,
and $a,b$ is an adjoint representation index.}:
\begin{eqnarray}
\gamma^{(1)ai}_{bj}&=&\frac{1}{\lf}\left\{2C_1(\lambda^2-g^2)\delta_{ij}+
\frac{N^2-4}{N}C_2^3h^{(2)}_{ij}\right\}\delta_{ab},\nonumber\\
\beta^{(1)}_{\lambda}&=&\frac{\lambda}{\lf}\left\{6C_1(\lambda^2-g^2)
    +\frac{N^2-4}{N}C_2^3\Tr(h^{(2)})\right\},\\
\beta^{(1)}_{h_{ijk}}&=&\frac{1}{\lf}\left\{6C_1(\lambda^2-g^2)h_{ijk}
  +\frac{N^2-4}{N}C_2^3h^{(3)}_{ijk}\right\}\nonumber,
\end{eqnarray}
where we defined:
\begin{eqnarray}
 h^{(3)}_{ijk}&\equiv&h^*_{plm}(h_{ijp}h_{klm}+h_{kjp}h_{ilm}+h_{ikp}h_{jlm})\nonumber\\
 h^{(2)}_{ij}&\equiv& h_{ilm}h^*_{jlm}.
 \end{eqnarray}

The equations simplify if we rescale the coupling constants:
\begin{equation}
g\to\frac{\sqrt{C_1}}{4\pi}g, \quad
\lambda\to\frac{\sqrt{C_1}}{4\pi}\lambda  \quad\hbox{and}\quad
h_{ijk}\to\frac{\sqrt{C_2^3}}{4\pi}\sqrt{\frac{N^2-4}{N}}h_{ijk}.
\end{equation}
The $\beta$-functions become:
\begin{equation}
\beta_g=-\frac{2g^3}{1-2g^2}\Tr(\gamma), \quad
\beta_{\lambda}=\lambda \Tr(\gamma).
\end{equation}
Here the trace is taken only over the $SU(3)$ indices and not over
gauge indices.

From these $\beta$-functions we can obtain a
differential equation:
\begin{equation}
  -\frac{1}{2g^3}dg+\frac{1}{g}dg=\frac{d\lambda}{\lambda}.
\end{equation}
This can be easily solved to give:
\begin{equation}
\frac{\lambda}{\lambda_0}=\frac{ge^{\frac{1}{4g^2}}}{g_0e^{\frac{1}{4g_0^2}}}.
\end{equation}
This result means that the RG flow lines in the $\lambda-g$ plane
are exactly known (to the extent that we can count on the NSVZ
$\beta$-function). It is easy to convince oneself that there is no
line with both couplings going to zero in the UV, except the
trivial case when one of the couplings is constantly zero. This
implies that there is no choice of coupling constants for which
this theory is asymptotically free.

 In order to have a fixed point we have to satisfy $\Tr(\gamma)=0$, which
implies at one loop that:
\begin{equation} \label{condFix}
  \Tr(h^{(2)})=-6(\lambda^2-g^2),
\end{equation}
and we can substitute this into $\beta_{h_{ijk}}$ to get another
condition:
\begin{equation}
\Tr(h^{(2)})h_{ijk}=h^{(3)}_{ijk}. \label{condFix_h}
\end{equation}
By multiplying (\ref{condFix_h}) on both sides by $h^*_{ijk}$ we
get:
\begin{eqnarray}
  3\Tr((h^{(2)})^2)&=&(\Tr(h^{(2)}))^2\label{cond1},
\end{eqnarray}
which implies $h^{(2)}_{ij}=\alpha^2\delta_{ij}$, and then $\gamma$
is proportional to the identity matrix. One can show that this implies that
we turn on
only the $\lambda$, $h_{111}=h_{222}=h_{333}$ and $h_{123}$ couplings (or
their $SU(3)$ rotations), and that $\alpha^2 =
\frac{1}{3}\sum_{i,j,k}{|h_{ijk}|^2}$.

The fixed points we found are IR stable fixed points,
since we have:
\begin{equation}
  \Tr(\gamma)=3(2(\lambda^2-g^2)+\alpha^2),
\end{equation}
and the condition for a fixed point is $\Tr(\gamma)=0$. From the
$\beta$-functions we calculated we see that if we increase one of
the couplings $\lambda$ or $h_{ijk}$, $\Tr(\gamma)$ becomes positive thus
decreasing these couplings and increasing the gauge coupling in
IR, till we get again zero. A similar behavior arises if we decrease the
couplings. Thus, we conclude that in the weak coupling limit
all fixed points that exist imply diagonal $\gamma$ and are IR
stable.

The general analysis above implies that the three dimensional
surface of fixed points should persist also at strong coupling. The
gravitational dual of these \textit{exactly} marginal deformations will be
discussed in \cite{a:AKS} (it was also discussed in \cite{a:petrini}).

\section{Orbifold Theories}\label{sec_orb}

It is possible to reduce the number of supersymmetries of the $d=4$
 ${\cal N}=4$ SYM, which we discussed in the previous section, via the
 orbifolding
 procedure, by looking at the theories arising from D3-branes at
 orbifold singularities \cite{a:quiver,a:quiver2,a:SilKach}.

 We will look at $N$ coincident D3-branes at the $\mathbb{Z}_k$
 orbifold singularity of a $\mathbb{C}^3/\mathbb{Z}_k$ space.
Denoting the $\mathbb{C}^3$ coordinates transverse to the D3-branes by $Z^l$,
the orbifold group acts on them as
\begin{eqnarray}
   Z^l\to \omega^{a_l}Z^l,
\end{eqnarray}
where $\omega\equiv e^{2\pi i\over k}$ and $(a_1,a_2,a_3)$ is a
triple of integers\footnote{The vector $\vec{a}$ has to satisfy
$\sum_i a_i=0$ (mod $k$) in order that the orbifold action will be
part of $SU(3)$ and not $U(3)$, which is the condition for
preserving supersymmetry. If we choose all the components to be
non zero we have $\mathbb{Z}_k\subset SU(3)$ thus leaving us with
one supersymmetry. If we choose one of the components zero then we
can have $\vec{a}=(n,0,-n)$ (mod $k$). This case is equivalent to
the $(1,0,-1)$ case, and in this case $\mathbb{Z}_k\subset SU(2)$
and we have ${\cal N}=2$ supersymmetry. So there is only one
choice giving an ${\cal N}=2$ theory here. If two components of
$\vec{a}$ vanish the remaining one must also vanish, giving the
${\cal N}=4$ case. }.

 To see
 how the orbifold acts on the D-branes we put $kN$
 D3-branes on the covering space and group them in $N$ sets of
 $k$ branes. We put each set of D-branes in the regular
 representation of $\mathbb{Z}_k$,
which is the direct sum of $k$ one dimensional irreducible
representations parameterized by the integer $n$, given by
$\omega^n$ (with $\omega$
 defined above). We denote each brane by a pair of indices,
 $i=0,\cdots,N-1$ and $I=0,\cdots,k-1$.

The bosonic fields on the D3-branes are a gauge field and three
complex scalars $Q_i$, sitting in chiral multiplets, whose eigenvalues
label the $Z^i$ positions of the branes.
The projection on
 the gauge fields is given by (we write the adjoint fields in double
 index notation, with the upper index in the fundamental
 representation and the lower in the anti-fundamental):
 \begin{eqnarray}
   A^{I,i}_{J,j}=\omega^{I-J}A^{I,i}_{J,j},
 \end{eqnarray}
and the projection on the chiral multiplets
 (which are related to the 6 transverse directions) is:
 \begin{eqnarray}
   (Q_l)^{I,i}_{J,j}=\omega^{I-J+a_{l}}(Q_l)^{I,i}_{J,j}.
 \end{eqnarray}
This projection comes from acting on the Chan-Paton indices as
well as on the space-time index.

The gauge fields which survive the
 projection are $A^{I,i}_{J,j}$ where both indices lie in
 the same irreducible representation of $\mathbb{Z}_k$, $I=J$,
giving a total of $k$ copies of
 $U(N)$ \footnote{We will treat the gauge groups as $SU(N)$ rather than
 $U(N)$ from here on, since one $U(1)$ factor is decoupled and the
 others are free in the IR.}. The
 matter content of the theory consists of chiral superfields
in the following representations:
 \begin{eqnarray}
\oplus_{I=0}^{k-1}(1,1...1,N_{(I)},1...1,\bar N_{(I+a_1)},1....1,1),\nonumber\\
\oplus_{I=0}^{k-1}(1,1...1,N_{(I)},1...1,\bar N_{(I+a_2)},1....1,1),\nonumber\\
\oplus_{I=0}^{k-1}(1,1...1,N_{(I)},1...1,\bar
  N_{(I+a_3)},1....1,1),
\end{eqnarray}
where $N_{(I)}$ labels the $I$'th $SU(N)$ factor, except when $a_I=0$
in which case we get a chiral superfield in the adjoint of $U(N)^k$.
We will denote these fields by $Q_l^I$ where $I\in
(0,...,k-1)$, $l\in(1,2,3)$. The index $I$ denotes the index of the
$SU(N)$ group of which the field is in the fundamental (or adjoint)
representation.

The matter content of this theory can be summarized in a
``quiver" diagram, where vertices represent the gauge groups, and
oriented lines represent chiral multiplets in the fundamental of
the group to which they point and the antifundamental of the
second group.

\begin{figure}[htbp]
\centerline{\epsfxsize=4in\epsfbox{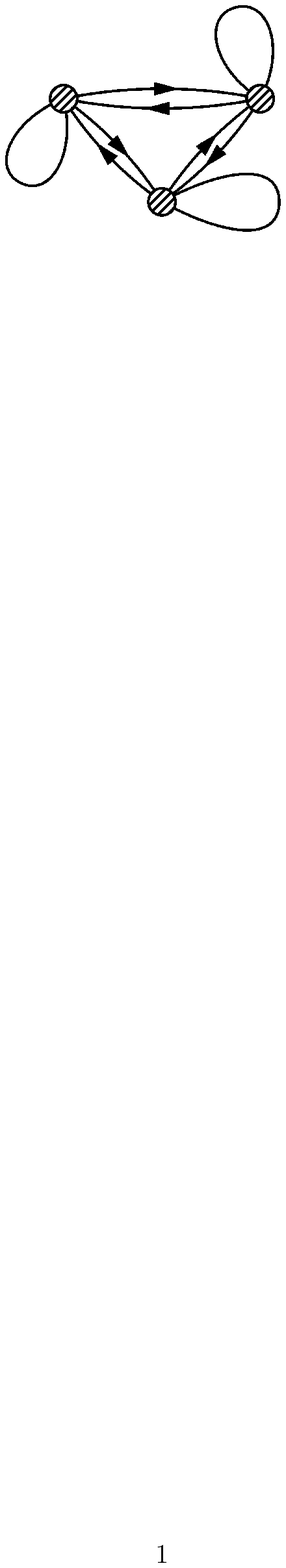}}
\caption{$\mathbb{C}^3/\mathbb{Z}_3$ (1,-1,0) quiver diagram}
\label{quiv}
\end{figure}

In this paper we won't be interested in particular in the
superpotential coming from
the orbifold theory, but rather in the most general superpotential
with this matter content which is classically marginal.
Generally, the only possible marginal
superpotential\footnote{In the quiver diagram, the possible
superpotentials (for $SU(N > 3)$) correspond to oriented triangles.}
is of the form:
\begin{eqnarray}\label{sup_pot}
  W=h^I_{lmn} \Tr(Q^I_lQ^{I+a_l}_mQ^{I+a_l+a_m}_n),
\end{eqnarray}
with $a_l+a_m+a_n=0\ ({\rm{mod}}\ k)$. The
definition of the couplings $ h^I_{lmn}$ in this way is redundant,
since
\begin{eqnarray}\label{redu}
   h^I_{lmn}= h^{I+a_l}_{mnl}= h^{I+a_l+a_m}_{nlm}.
\end{eqnarray}

Obviously, for a general choice of $k$ and of $\vec{a}$,
the only possibility is to take
$(l,m,n)$ to be some permutation of $(1,2,3)$.
In some special cases additional superpotentials are possible. For $N
> 3$ these cases are :
\begin{itemize}
\item ${\cal N}=1$ SUSY
\begin{itemize}
 \item general $k$, $\vec{a}=(a,a,-2a)$,
 \item $k=3k'$,    $\vec{a}=(a,{k\over
3}-a,-{k\over3})$,
\item $k=3k'$, $\vec{a}=({k\over 3},{k\over
 3},{k\over3})$,
\item $k=6k'$, $\vec{a}=({k\over
6},{k\over 6},{2k\over3})$.
\end{itemize}
\item ${\cal N}=2$ SUSY
\begin{itemize}
\item $k=3$, $\vec{a}=(1,-1,0)$.
\end{itemize}
\end{itemize}

When $a_1,a_2$ and $k$ have a common divisor $J$ larger than one,
then the theory splits into $J$ copies of the
$\mathbb{Z}_{k\over J}$, ${1\over J}\vec{a}$ theory. In
particular
there is no meaning to discussing the $\vec{a}=({k\over 3},{k\over
3},{k\over3})$, $\vec{a}=({k\over
6},{k\over 6},{2k\over3})$ theories for general $k$, they are all
equivalent to the ones with $k=3,6$ respectively. We will
assume that $a_1,a_2$ and $k$ have no non-trivial common
divisor.

When the gauge group is $SU(3)^k$ there is another possible set of
superpotentials. If $Q^I_i,Q^I_j,Q^I_p$ are bifundamentals of the same
two groups, we can add the following
marginal operator:
\begin{eqnarray}\label{N3}
 W = {\rho^I_{ijp}\over
 3!}\epsilon_{lmn}\epsilon^{abc}(Q^I_i)^{l}_a(Q^I_j)^{m}_b(Q^I_p)^{n}_c.
\end{eqnarray}

We will start by treating the most general case in some detail,
and then discuss briefly the special cases. Only fields in the
same representation can mix under renormalization,
 so we can write the gamma matrix as
 $\gamma^I_{lm}$, where $\gamma^I_{lm}$ can be non
 vanishing only if $a_l=a_m$. In all the orbifold theories the 1-loop
 beta functions vanish, and the NSVZ formula gives
 \begin{eqnarray}\label{betta1}
   \beta_{g_I}\propto
   \Tr(\gamma^I)+\gamma^{I-a_1}_{11}+\gamma^{I-a_2}_{22}+\gamma^{I-a_3}_{33}.
 \end{eqnarray}
For the superpotential couplings we have the usual expression coming from the
 general formula (\ref{betaY}),
 \begin{eqnarray}\label{betta2}
 \beta_{h^I_{lmn}}\propto
 h^I_{pmn}\gamma^I_{pl}+h^I_{lpn}\gamma^{I+a_l}_{pm}+h^I_{lmp}\gamma^{I+a_l+a_m}_{pn}.
 \end{eqnarray}

\subsection{The General Case}

As we mentioned at the beginning of this section, in the most general
case $h^I_{ijk}$ (from (\ref{sup_pot})) can appear only if
$(i,j,k)$ is some permutation of $(1,2,3)$. The redundancy
condition gives
\begin{eqnarray}
   h^I_{123}&=&h^{I+a_1}_{231}=h^{I+a_1+a_2}_{312},\nonumber\\
   h^I_{132}&=&h^{I+a_1}_{321}=h^{I+a_1+a_3}_{213},
\end{eqnarray}
so we actually have only $2\times k$ independent superpotential
 couplings here, $h_I \equiv h^I_{123}$ and $h'_I \equiv h^I_{132}$.

We start with a general analysis of the expected dimension of the
space of \textit{exactly} marginal deformations, along the lines of
\cite{a:Leigh}. From the general
 $\beta$-functions (\ref{betta1}), (\ref{betta2}), we obtain,
using the fact that in this
 case there is no mixing between the fields,
\begin{eqnarray}\label{an_dim}
   \beta_{h_I}&\propto& \gamma_1^I+\gamma_2^{I+a_1}+
\gamma_3^{I+a_1+a_2},\nonumber\\
   \beta_{h'_I}&\propto&
   \gamma_1^I+\gamma_3^{I+a_1}+\gamma_2^{I+a_1+a_3},\\
   \beta_{g_I}&\propto&
   \gamma_1^I+\gamma_2^I+\gamma_3^I+\gamma_1^{I+a_2+a_3}+\gamma_2^{I+a_1+a_3}+\gamma_3^{I+a_1+a_2}.\nonumber
 \end{eqnarray}
Naively the vanishing of (\ref{an_dim}) gives
$3k$ conditions for $3k$ couplings so we
don't expect any \textit{exactly} marginal directions. However, there are some
relations between the $\beta$-functions.

Let us denote the largest common divisor of $k$ and  $a_i$ by
$\alpha_i$, and define ${\cal S}_{a_i}^J$ to be the set of
indices $(J,J+a_i,J+2a_i,\cdots)$. We find that
\begin{eqnarray}
  \sum_{I\in{\cal S}_{a_1}^J}{\beta_{h_I}\over h_I}&\propto& \sum_{I\in{\cal
      S}_{a_1}^J}(\gamma_1^I+\gamma_2^{I})+\sum_{I\in{\cal S}_{a_1}^{J+a_2}}
\gamma_3^{I+a_2},\nonumber\\
   \sum_{I\in{\cal S}_{a_1}^J}{\beta_{h'_I}\over h'_I}&\propto&
   \sum_{I\in{\cal S}_{a_1}^J}(\gamma_1^I+\gamma_3^{I})+
\sum_{I\in{\cal S}_{a_1}^{J+a_1+a_3}}\gamma_2^{I+a_1+a_3},\\
\sum_{I\in{\cal S}_{a_1}^J}{\beta_{g_I}\over f(g_I)}&\propto& \sum_{I\in{\cal
      S}_{a_1}^J}(2\gamma_1^I+\gamma_2^I+\gamma_3^I)+
\sum_{I\in{\cal S}_{a_1}^{J+a_1+a_3}}\gamma_2^{I+a_1+a_3}+\sum_{I\in{\cal
      S}_{a_1}^{J+a_2}}\gamma_3^{I+a_2},\nonumber
 \end{eqnarray}
where $f(g)\equiv{1\over 16\pi^2}{2g^3C_1\over 1-{2C_1g^2\over
16\pi^2}}$. Thus, we have $\sum_{I\in{\cal
S}_{a_1}^J}{\beta_{g_I}\over f(g_I)}\propto
 \sum_{I\in{\cal S}_{a_1}^J}{\beta_{h_I}\over h_I}+ \sum_{I\in{\cal
 S}_{a_1}^J}{\beta_{h'_I}\over h'_I}$, and our system of linear equations
 is dependent. The number of such dependencies is obviously
      $\alpha_1$, since
there are ${k\over\alpha_1}$ elements in ${\cal S}_{a_1}^J$. We
can do the same procedure for $a_2$ and $a_3$, finding
$\sum_{I\in{\cal S}_{a_2}^J}{\beta_{g_I}\over f(g_I)}\propto
 \sum_{I\in{\cal S}_{a_2}^J}{\beta_{h_I}\over h_I}+ \sum_{I\in{\cal
 S}_{a_2}^J}{\beta_{h'_I}\over h'_I}$, $\sum_{I\in{\cal S}_{a_3}^J}{\beta_{g_I}\over f(g_I)}\propto
 \sum_{I\in{\cal S}_{a_3}^J}{\beta_{h_I}\over h_I}+ \sum_{I\in{\cal
 S}_{a_3}^J}{\beta_{h'_I}\over h'_I}$. These three relations are
 not completely independent. By summing over $J$ each of
      the
 three relations we get the same constraint. Thus, we find from here
 $(\sum^3_{i=1}\alpha_i-2)$ linear relations between the beta functions.

The $\beta_{h_I}$ and $\beta_{h'_I}$ are also not
 completely independent: we have
$\sum_I{\beta_{h_I}\over h_I}=\sum_I{\beta_{h'_I}\over
 h'_I}$, which gives another relation.
Thus, all in all we
  have $(\sum^3_{i=1}\alpha_i-1)$ linear relations
 between the $\beta$-functions.

We have $k$ gauge
  couplings,  $k$ $h_I$'s, and $k$ $h'_I$'s, for a total of $3k$
  parameters. We have
  $3k-(\sum^3_{i=1}\alpha_i-1)$ independent
  equations, so we expect to find an
  $(\sum^3_{i=1}\alpha_i-1)$ dimensional
  manifold of fixed points.
In the generic case of $\alpha_i=1$
we find two \textit{exactly} marginal deformations. This is what we expect
from the orbifold relations, at least for large $N$, as we
will discuss in section 4.

We also found that some linear combinations of anomalous dimensions do
  not appear in (\ref{an_dim}), so the anomalous
  dimensions do not have to vanish on the fixed surface, unlike the
  ${\cal N}=4$ case.
Let us find explicitly the $(\sum_{i=1}^3 \alpha_i-1)$-dimensional
  space of possible values for the anomalous
  dimensions. From the vanishing of (\ref{an_dim}) we get:
 \begin{eqnarray}
  -\gamma^I_3&=&\gamma_2^{I+a_1+a_3}+\gamma_1^{I+a_3},\nonumber\\
   -\gamma^I_3&=&\gamma_1^{I+a_2+a_3}+\gamma_2^{I+a_3},\\
    -\gamma^I_3-\gamma^{I+a_1+a_2}_3&=&\gamma_2^{I+a_1+a_3}+
\gamma_1^{I+a_2+a_3}+\gamma^I_1+\gamma^I_2,\nonumber
  \end{eqnarray}
leading to
  \begin{eqnarray}
\gamma_2^{I+a_1+a_3}+\gamma_1^{I+a_3}&=&\gamma_2^{I+a_3}+
\gamma_1^{I+a_2+a_3},\nonumber\\
\gamma_2^{I+a_1+a_3}+\gamma_1^{I}&=&\gamma_2^{I+a_3}+\gamma_1^{I+a_2},
  \end{eqnarray}
and finally to
  \begin{eqnarray}
     \gamma_1^{I+a_3}-\gamma_1^{I+a_2+a_3}=\gamma_1^{I}-\gamma_1^{I+a_2}.
  \end{eqnarray}
We see that if we define $\gamma_1^{I}-\gamma_1^{I+a_2}\equiv
  K_I$ then $K_I=K_{I+a_3}$, so we have $\alpha_3$ independent $K_I$'s,
which satisfy $\sum_{I\in{\cal S}_{a_2}^J}K_I=0$.

Since $a_2,a_3$ and $k$ have no common divisor, we find that if
$\sum_{I\in{\cal
  S}_{a_2}^J}K_I=0$ for some $J$ then it is true for any
  $J$. Similarly, we find that
$\sum_{I\in{\cal S}_{a_1}^J}K_I=0$.
Thus, we have one constraint on $\alpha_3$  $K_I$'s. Since
  \begin{eqnarray}\label{rels_gamma}
   \gamma_1^{I}-K_I&=&\gamma_1^{I+a_2},\nonumber\\
    \gamma_2^{I}-K_I&=&\gamma_2^{I+a_1},\\
     -\gamma_3^{I}&=&\gamma_1^{I+a_2+a_3}+\gamma_2^{I+a_3},\nonumber
  \end{eqnarray}
we conclude that we have $\alpha_2$ independent
$\gamma_1^I$'s, $\alpha_1$ independent $\gamma_2^I$'s and
$\alpha_3-1$ independent $K_I$'s. Thus, we can have
$(\sum^3_{i=1}\alpha_i-1)$ independent $\gamma$-functions as expected.
The naive expectation is that these $\gamma$ functions will not vanish
at generic points on the surface of \textit{exactly} marginal deformations.

\subsubsection{Perturbative Calculations}

Let us now look at the perturbative conditions for vanishing
$\beta$-functions. At one loop we have\footnote{We define ${\cal
B}={1\over16\pi^2}{N^2-1\over N}$, ${\cal
A}={1\over16\pi^2}{N\over 2}$.}
\begin{eqnarray}
   \gamma_{Q^I_1}&=&{\cal A}(|h^{I}_{123}|^2+|h^{I}_{132}|^2)-{\cal
   B}(g_{I}^2+g_{I+a_1}^2)=\gamma^I_1,\nonumber\\
 \gamma_{Q^I_2}&=&{\cal A}(|h^{I}_{231}|^2+|h^{I}_{213}|^2)-{\cal
   B}(g_{I}^2+g_{I+a_2}^2)=\gamma^I_2,\nonumber\\
 \gamma_{Q^I_3}&=&{\cal A}(|h^{I}_{312}|^2+|h^{I}_{321}|^2)-{\cal
   B}(g_{I}^2+g_{I+a_3}^2)=\gamma^I_3,
\end{eqnarray}
or
\begin{eqnarray}\label{1ll}
  \gamma_{Q^I_1}&=&{\cal A}(|h_{I}|^2+|h'_{I}|^2)-{\cal
   B}(g_{I}^2+g_{I+a_1}^2)=\gamma^I_1,\nonumber\\
 \gamma_{Q^I_2}&=&{\cal A}(|h_{I-a_1}|^2+|h'_{I+a_2}|^2)-{\cal
   B}(g_{I}^2+g_{I+a_2}^2)=\gamma^I_2,\nonumber\\
 \gamma_{Q^I_3}&=&{\cal A}(|h_{I+a_3}|^2+|h'_{I-a_1}|^2)-{\cal
   B}(g_{I}^2+g_{I+a_3}^2)=\gamma^I_3.
\end{eqnarray}

Defining
 \begin{eqnarray}
  A_{I+a_1}&\equiv&{\cal A}|h_I|^2-{\cal B}g_{I+a_1}^2,\quad\quad B_I\equiv
  {\cal A}|h'_I|^2-{\cal B}g_I^2,\nonumber\\
  C_I&\equiv& {\cal B}(
  g_{I-a_1}^2+g_{I-a_2}^2-g_I^2-g_{I+a_3}^2),
 \end{eqnarray}
we have using (\ref{rels_gamma}):
\begin{eqnarray}\label{one_l}
 \gamma_{Q_1^I}&=&A_{I+a_1}+B_I=\gamma_1^I,\nonumber\\
 \gamma_{Q_2^I}&=&A_{I}+B_{I+a_2}=\gamma_2^I,\nonumber\\
 \gamma_{Q_3^{I+a_1+a_2}}&=&A_{I+a_1}+B_{I+a_2}+C_{I+a_1+a_2}=
-(\gamma_1^I+\gamma_2^I-K_I).
\end{eqnarray}

By subtracting the first equation from the third and summing over
${\cal S}_{a_2}^J$ we get zero since $\sum_{I\in{\cal S}_{a_2}^J}
C_I=0$, and thus we find $\sum_{I\in{\cal
S}_{a_2}^J}(2\gamma_1^I+\gamma_2^I-K_I)=\sum_{I\in{\cal
S}_{a_2}^J}(2\gamma_1^I+\gamma_2^I)=0$, and further by using
(\ref{rels_gamma}) we find
\begin{equation}\label{one_o}
-{2k\over
\alpha_2}\gamma_1^J=\sum_{I\in{\cal
S}_{a_2}^J}\gamma_2^I-2\sum_{j=0}^{{k\over\alpha_2}-1}({k\over\alpha_2}-
1-j)K_{I+ja_2}.
\end{equation}

 By subtracting the second equation from the
third and summing over ${\cal S}_{a_1}^J$ we again find zero, so
$\sum_{I\in{\cal
S}_{a_1}^J}(\gamma_1^I+2\gamma_2^I-K_I)=\sum_{I\in{\cal
S}_{a_1}^J}(\gamma_1^I+2\gamma_2^I)=0$, and further by using
(\ref{rels_gamma}) we find
\begin{equation}\label{two_o}
-{2k\over
\alpha_1}\gamma_2^J=\sum_{I\in{\cal
S}_{a_1}^J}\gamma_1^I-2\sum_{j=0}^{{k\over\alpha_1}-1}({k\over\alpha_1}-
1-j)K_{I+ja_1}.
\end{equation}

Plugging (\ref{two_o}) into (\ref{one_o}) we
find\footnote{We define ${\cal
T}^J_{a_i,a_j}\equiv\{I|I=J+n_1a_i+n_2a_j;
n_1,n_2\in\mathbb{N}\}$.} :
\begin{eqnarray}\label{condic}
{4k^2\over \alpha_1\alpha_2}\gamma_1^J&=&\sum_{I\in{\cal
T}_{a_1,a_2}^J}\gamma_1^I-2\sum_{l,j}({k\over\alpha_1}-1-j)K_{J+ja_1+la_2}
+{4k\over\alpha_1}\sum_{j=0}^{{k\over\alpha_2}-1}({k\over\alpha_2}-1-j)
K_{J+ja_2}=\nonumber\\
&=&\sum_{I\in{\cal
T}_{a_1,a_2}^J}\gamma_1^I-{4k\over\alpha_1}\sum_{j=0}^{{k\over\alpha_2}-1}
jK_{J+ja_2},
\end{eqnarray}
where in the last line we used $\sum_{I\in{\cal
T}^J_{a_1,a_2}}K_I=0$. From the first two equations in
(\ref{one_l}) it is clear that
\begin{equation}\label{gamma_rel}
\sum_{I\in{\cal
S}_{a_1+a_2}^{J+a_2}}\gamma_1^{I+a_2}=\sum_{I\in{\cal
S}_{a_1+a_2}^J}\gamma_2^{I},
\end{equation}
and that $\sum_I\gamma_1^I=\sum_I\gamma_2^I$, leading to
$\sum_I\gamma_1^I=\sum_I\gamma_2^I=0$. From here and from
(\ref{condic}) we see that $\gamma_{1}^I=-{\alpha_2\over
k}\sum_{j=0}^{{k\over\alpha_2}-1}jK_{I+ja_2}$ and
$\gamma_{2}^I=-{\alpha_1\over
k}\sum_{j=0}^{{k\over\alpha_1}-1}jK_{I+ja_1}$. The periodicity of
$K_I$ now implies that $\gamma_1^I=\gamma_1^{I+a_1+a_2}$ and
$\gamma_2^I = \gamma_2^{I+a_1+a_2}$, and from (\ref{gamma_rel}) we
get $\gamma_1^{I+a_2} = \gamma_2^I$. Now, from (\ref{rels_gamma})
we get:
\begin{equation}
\gamma_1^{I+a_1+a_2}=\gamma_2^{I+a_1}
=\gamma_2^I-K_I=\gamma_1^{I+a_2}-K_I= \gamma_1^I-2K_I,
\end{equation}
so all $K_I$ have to vanish. Using our equations this implies that
all the $\gamma^I_{1,2,3}$
have to vanish. So, at one loop order we cannot turn on any non
vanishing anomalous dimensions.

 From the first and second equations in (\ref{one_l}) we now see that
 $A_I=A_{I+a_1+a_2}$, so we can parameterize our solution by
 $\alpha_3$ $A_I$'s. The $B_I$'s are obtained from the $A_I$'s using
 (\ref{one_l}). Using the second and third equations we have
 $A_{I+a_1}-A_I=-C_{I+a_1+a_2}$, giving the $C_I$'s. Note that from the
 definition of $C_I$ we have $\sum_{I\in{\cal
S}_{a_2}^J}C_I,\sum_{I\in{\cal S}_{a_1}^J}C_I=0$, thus there are
only $k+1-\alpha_1-\alpha_2$ independent $C_I$'s (the shift by one
is because $\sum_IC_I=0$ follows from both constraints). From the
$C_I$'s we get linear equations on $k$ gauge couplings squared,
whose
 space of solutions is
$\alpha_1+\alpha_2-1$-dimensional. The $h_I$'s and
$(h'_I)$'s can be obtained from the $B_I$'s, the $A_I$'s and the gauge
couplings.

Thus, to summarize, we have $\sum_i\alpha_i-1$ parameters for our
solution, as expected. The solution obtained looks non generic, since
in the general analysis we expected to find a
$(\sum_i\alpha_i-1)$-dimensional manifold of fixed points due to the
possibility of turning on non-zero anomalous dimensions. However, we
see that at one loop the anomalous dimensions are forced to vanish, and
nevertheless we find the expected dimensionality of the manifold of fixed
points. The one-loop solution can be extended to all orders in
perturbation theory, as we will illustrate below in the
${\cal N}=2$ example. Generally, the anomalous dimensions may be
turned on in this solution at higher loop orders.

\subsubsection{${\cal N}=2$ example}\label{sec_N_2}

We will illustrate the results of the previous section by the
${\cal N}=2$ example, which is the
$\mathbb{C}^3/\mathbb{Z}_k$ $(1,-1,0)$ orbifold. We denote
$Q_I \equiv Q_1^I$, $\tilde Q_I \equiv Q_2^{I+1}$ and $\Phi_I \equiv Q_3^I$.

The possible superpotentials here are:
\begin{eqnarray}\label{pts}
  W_1&=&{1 \over 6}\Tr(\alpha_I\tilde Q_I\Phi_I Q_I+
\delta_I Q_I\Phi_{I+1}\tilde Q_I),\nonumber\\
  W_2&=&{1 \over 6}h_I\Tr(\Phi_I\Phi_I\Phi_I).
\end{eqnarray}
The superpotential $W_2$ is specific to the ${\cal N}=2$ case
where $a_3=0$ so we will deal with it at the end of this
subsection. Setting $h_I$ to zero, the $\beta$-functions are :
\begin{eqnarray}\label{betas}
  \beta_{g_I}&=&-{2g_I^3\over 16\pi^2}{N\over
  1-{2Ng_I^2\over16\pi^2}}
({1\over2}(\gamma_{Q_I}+\gamma_{\tilde Q_I}+\gamma_{Q_{I-1}}+
\gamma_{\tilde Q_{I-1}})+\gamma_{\Phi_I}),\nonumber\\
  \beta_{\alpha_I}&=&\alpha_I(\gamma_{Q_I}+\gamma_{\tilde Q_I}+
\gamma_{\Phi_I}),\\
  \beta_{\delta_I}&=&\delta_I(\gamma_{Q_I}+\gamma_{\tilde
  Q_I}+\gamma_{\Phi_{I+1}}).\nonumber
\end{eqnarray}

Because of the symmetry of all interactions we have
$\gamma_{Q_I}=\gamma_{\tilde Q_I}$. Equating the $\beta$-functions
to zero we obtain that  $\forall I:\gamma_{\Phi_I}\equiv \gamma$,
and also all the $\gamma_{Q_I}$'s have to be equal and equal to
$-{1\over2}\gamma$. So, we have here $3k$ couplings and one
possible independent anomalous dimension $\gamma$. Since a priori
we have $2k$ different anomalous dimensions, we have $2k-1$ equations, and
we expect to find a
$(k+1)$-dimensional manifold of fixed points.

Next, we do the perturbative analysis. The one-loop calculation
gives:
\begin{eqnarray}
  \gamma_{\Phi_I}&=&{1 \over 16\pi^2}{N\over 4}
((|\delta_{I-1}|^2+|\alpha_I|^2)-8g_I^2),\\
  \gamma_{Q_I}&=&{1 \over 16\pi^2}{N^2-1\over4N}\left\{
(|\delta_I|^2+|\alpha_I|^2)-4(g_I^2+g_{I+1}^2)\right\}.\nonumber
\end{eqnarray}
Defining $B_I\equiv |\delta_{I-1}|^2-4g_I^2$,
$A_I\equiv|\alpha_I|^2-4g_I^2$ and $16\pi^2{4N\over
N^2-1}\gamma\to\gamma$, the requirement of vanishing
 $\beta$-functions becomes :
\begin{eqnarray}\label{one_loop_sol}
  B_I+A_I&=&{N^2-1\over N^2}\gamma,\\
  B_{I+1}+A_I&=&-{1\over2}\gamma.\nonumber
\end{eqnarray}
By subtracting the first line from the second and summing over
$I$, we find that $\gamma=0$. Thus, again we find that at one loop
precision the $\gamma$ parameter has to vanish. As we will see
later this is not necessarily true for higher loop calculations.

The case of vanishing $\gamma$ is the case of
vanishing anomalous dimensions.
 We see that in this case the condition for having zero
$\beta$-functions is that for all $I$,
$B_I= X=-A_I$ for some number $X$ which is a
parameter. Thus, we find a family of solutions parameterized
by $X$ and the gauge couplings, with
\begin{eqnarray}\label{sol}
  |\delta_{I-1}|^2&=&X+4g_I^2,\\
  |\alpha_I|^2&=&4g_I^2-X.\nonumber
\end{eqnarray}
We see that the parameter $X$ is constrained to the range
$-\min_{I}\left\{4g_I^2\right\}\leq X\leq
\min_{I}\left\{4g_I^2\right\}$. The case $X=0$ is the case of
${\cal N}=2$ SUSY.

To summarize, we find a $(k+1)$-dimensional space of solutions. We
expected a $(k+1)$-dimensional manifold from the general
analysis, but the $(+1)$ was due to the $\gamma$ parameter.
At one loop we find that $\gamma=0$ but nevertheless we have a
$(k+1)$-dimensional space of solutions.

A natural question is whether the vanishing of $\gamma$ extends to
higher loops, and whether we can extend our solution, parameterized by
the gauge couplings and the parameter $X$, to
higher loops.
We will prove that the non-vanishing $X$ solution does not
 disappear at higher loops. First we will represent a general solution as a
 function of the gauge couplings and the $X$ parameter. The
 procedure we use here is similar to the coupling constant
 reduction procedure described in \cite{a:Reduct}.

 The most general solution for $\alpha_I$ and
  $\delta_I$ depending on our parameters $X$ and $g_I$, consistent with the one
  loop analysis
  and with the ${\cal N}=2$ case (which is known to be exactly
  conformal for any gauge couplings), is of the form
\begin{eqnarray}\label{gen_sols}
  |\delta_{I-1}|^2&=&4g_I^2+X(1+\sum_{m,j,l_s}a^{(I)m}_{l_1...l_j}
X^mg_{l_1}^2...g_{l_j}^2),\\
  |\alpha_I|^2&=&4g_I^2-X(1+\sum_{m,j,l_s}b^{(I)m}_{l_1...l_j}X^m
g_{l_1}^2...g_{l_j}^2),\nonumber
\end{eqnarray}
where $a,b$ are some constants and $m+j>0$. We will construct the
  solution by an inductive process.
  Assume that we have computed the $a$'s and $b$'s in these solutions
  up to $(n-1)$'th order in $g^2$ and $X$, and look at the $n$'th order.
  First, we calculate the $\gamma_{Q_I}$ and $\gamma_{\Phi_I}$. We
  can write them as:
  \begin{eqnarray}\label{demand}
\gamma_{Q_I}^{(n)}=\gamma_{Q_I}^{(n)(1-loop)}+\gamma_{Q_I}^{(n)(2..n-loops)}\\
\gamma_{\Phi_I}^{(n)}=\gamma_{\Phi_I}^{(n)(1-loop)}+\gamma_{\Phi_I}^{(n)(2..n-loops)}.\nonumber
  \end{eqnarray}

 We define:
  \begin{eqnarray}
  \tilde B_I^{(n)}&\equiv&(\delta_{I-1}^2)^{(n)}=
X\cdot(\sum_{m,j,l_s,m+j=(n-1)}a^{(I)m}_{l_1...l_j}X^mg_{l_1}^2...g_{l_j}^2),\\
  \tilde A_I^{(n)}&\equiv&(\alpha_I^2)^{(n)}= -
  X\cdot(\sum_{m,j,l_s,m+j=(n-1)}b^{(I)m}_{l_1...l_j}X^mg_{l_1}^2...g_{l_j}^2).
\nonumber
\end{eqnarray}
The $\gamma_{Q_I}^{(1-loop)}$ and $\gamma_{\Phi_I}^{(1-loop)}$
have a special structure, giving:
\begin{eqnarray}
  \tilde B_{I+1}^{(n)}+\tilde A_I^{(n)}&=&\gamma_{Q_I}^{(n)(1-loop)},\\
  \tilde B_I^{(n)}+\tilde A_I^{(n)}&=&{N^2-1\over N^2}
\gamma_{\Phi_I}^{(n)(1-loop)}.\nonumber
\end{eqnarray}
Now, we parameterize the remaining contributions to the $\gamma$'s as:
\begin{eqnarray}
    \gamma_{Q_I}^{(n)(2..n-loops)}&\equiv&T_I^{(n)}+S_{I+1}^{(n)}-
{1\over2}\tilde\gamma^{(n)},\\
    {N^2-1\over N^2}\gamma_{\Phi_I}^{(n)(2..n-loops)}&\equiv&
T_I^{(n)}+S_I^{(n)}+{N^2-1\over
    N^2}\tilde\gamma^{(n)},\nonumber
\end{eqnarray}
where the different quantities are defined as :
\begin{eqnarray}
 -k({1\over2}+{N^2-1\over N^2})\tilde\gamma^{(n)}&\equiv&\sum_I{
\gamma_{Q_I}^{(n)(2..n-loops)}-{N^2-1\over
N^2}\gamma_{\Phi_I}^{(n)(2..n-loops)}},\nonumber\\
 \Delta X^{(n)}&=& S_1^{(n)} \equiv 0,\\
  \gamma_{Q_I}^{(n)(2..n-loops)}-{N^2-1\over
N^2}\gamma_{\Phi_I}^{(n)(2..n-loops)}&=&S^{(n)}_{I+1}-
S^{(n)}_I-({1\over2}+{N^2-1\over
N^2})\tilde\gamma^{(n)}.\nonumber
\end{eqnarray}
Note that $\Delta X^{(n)}$ is just a redefinition of X, so we can
set it to zero without any loss of generality, and the
$T_I^{(n)}$'s are automatically determined from above. We see that
the definitions above uniquely determine $S_I, T_I$ and
$\tilde\gamma$.

The crucial point is that in order to calculate the one loop
 contribution to the $n$'th order we use the $n$'th order components of
 (\ref{gen_sols}), while for two loops we use the (n-1)'th order of
 (\ref{gen_sols}), and so on. Thus, because we have already determined
 (\ref{gen_sols}) up to $n-1$'th order,
 $\gamma_{\Phi_I}^{(n)(2..n-loops)}$ and
 $\gamma_{Q_I}^{(n)(2..n-loops)}$
 depend only on already determined quantities. The yet
 undetermined quantities appear only at one loop.

From the $\beta$-function analysis we know that
 $\gamma_{Q_I}=-{1\over2}\gamma$ and $\gamma_{\Phi_I}=\gamma$.
 Thus, using the equations above we find
 \begin{eqnarray}
\tilde A_I^{(n)} &=& -T_I^{(n)},\\
\tilde B_I^{(n)} &=& -S_I^{(n)},\nonumber\\
 \tilde\gamma^{(n)} &=& \gamma^{(n)},\nonumber
\end{eqnarray}
where in the first two lines we are determining the $n$'th order
$a$'s and $b$'s, and in the third line we are computing the
$\gamma$ parameter. We see that it does not have to be zero at
higher loops.
This procedure is well defined and unique, and can be
 extended to any order in perturbation theory.
So, we have proven that
there exists a $(k+1)$-dimensional manifold of fixed points
parameterized by the gauge couplings and the $X$ parameter\footnote{
Equivalently, we can use $\gamma$ to parameterize the solution,
if it is non-zero starting
from some order in the perturbation series.}, at all orders of
perturbation theory.

The procedure described above for the ${\cal N}=2$ theory can be
repeated for any general orbifold theory. In the general case we
also find that the anomalous dimensions can not be turned on at
one loop order, but nevertheless the number of \textit{exactly}
marginal directions is as predicted from the general analysis.
The anomalous dimensions may be turned on at
higher orders of perturbation theory.

As mentioned above, in the specific ${\cal N}=2$ example we can
also turn on a deformation $W={1 \over
6}h_I\Tr(\Phi_I\Phi_I\Phi_I)$, which we now analyze. In
this case we are constrained to have $\gamma_{\Phi_I}=0$ or
equivalently $\gamma=0$. We have here $4k$ couplings ($k$ gauge
couplings, $\alpha_I$'s, $\delta_I$'s and $h_I$'s), and $2k$
constraints $\gamma_{\Phi_I}=\gamma_{Q_I}=0$, so we expect naively
to find a $2k$-dimensional manifold of fixed points.

At one-loop we get, defining $C_I\equiv{1\over  8}{N^2-4\over
N^2}|h_I|^2$,
\begin{eqnarray}\label{1_loop_C}
  B_I+A_I&=&-C_I,\\
  B_{I+1}+A_I&=&0.\nonumber
\end{eqnarray}
By subtracting the equations and summing over $I$ we get $\sum_I
C_I=0$, but this is impossible unless all $h_I$ vanish because
$C_I$ is positive definite.  So, we conclude that there are no
fixed points with non vanishing $h_I$ at one-loop.

 We now proceed in search of all loop solutions like we did above.
The general expressions for $\alpha_I$ and $\delta_I$ can now
depend also on the $h_I$'s, and we proceed as before:
  \begin{eqnarray}
\gamma_{Q_I}^{(n)}=\gamma_{Q_I}^{(n)(1-loop)}+
\gamma_{Q_I}^{(n)(2..n-loops)},\\
\gamma_{\Phi_I}^{(n)}=\gamma_{\Phi_I}^{(n)(1-loop)}+
\gamma_{\Phi_I}^{(n)(2..n-loops)},\nonumber
  \end{eqnarray}
where the $\gamma_{Q_I}^{(1-loop)}$ and
$\gamma_{\Phi_I}^{(1-loop)}$
  have a special structure,
\begin{eqnarray}
  \tilde B_{I+1}^{(n)}+\tilde A_I^{(n)}&=&\gamma_{Q_I}^{(n)(1-loop)},\\
  \tilde B_I^{(n)}+\tilde A_I^{(n)}+C_I^{(n)}&=&{N^2-1\over N^2}
\gamma_{\Phi_I}^{(n)(1-loop)}.\nonumber
\end{eqnarray}

Now, we parameterize the remaining contributions to the $\gamma$'s as
\begin{eqnarray}
    \gamma_{Q_I}^{(n)(2..n-loops)}&\equiv&T_I^{(n)}+S_{I+1}^{(n)},\\
    {N^2-1\over N^2}\gamma_{\Phi_I}^{(n)(2..n-loops)}&\equiv&
T_I^{(n)}+S_I^{(n)}-C_I^{(n)},\nonumber
\end{eqnarray}
where the different quantities are defined by
\begin{eqnarray}
 \sum_I C_I^{(n)}&=&\sum_I {\gamma_{Q_I}^{(n)(2..n-loops)}-{N^2-1\over
N^2}\gamma_{\Phi_I}^{(n)(2..n-loops)}},\nonumber\\
 \Delta X^{(n)}&=& S_1^{(n)}(\equiv 0),\\
  \gamma_{Q_I}^{(n)(2..n-loops)}-{N^2-1\over
N^2}\gamma_{\Phi_I}^{(n)(2..n-loops)}&=&S^{(n)}_{I+1}-S^{(n)}_I+C_I^{(n)}.
\nonumber
\end{eqnarray}
Again, $\Delta X^{(n)}$ is just a redefinition of  $X$, so we
can set it to zero without any loss of generality, the first equation
determines $\sum_I C_I$ and we can choose the individual $C_I$'s
arbitrarily (subject to this constraint), and then the
$T_I^{(n)}$'s are automatically determined from above.

The definitions above are well-defined, except for one caveat. The
first equation above cannot always be satisfied:
in the lowest
order where $C_I^{(n)}$ is not zero it has to be positive, so
at that order ${\hat \gamma} \equiv
\sum_I{\gamma_{Q_I}^{(n)(2..n-loops)}-{N^2-1\over
N^2}\gamma_{\Phi_I}^{(n)(2..n-loops)}}$ has to be positive. In
\cite{a:Thesis} we computed this combination $\hat\gamma$ up to
three loop order (without the $h_I$'s) and found that it vanished,
thus implying that no $h_I$'s can be turned on up to this order.

If at some higher order we find that $\hat\gamma$ is negative then
 there are no perturbative
solutions with non-zero $h_I$. On the other hand, if at
 the first order where it is non-zero it is positive, then
 we proceed like in the previous case to obtain a solution, by
 demanding that $\gamma_{Q_I}=\gamma_{\Phi_I}=0$, so that
 \begin{eqnarray}
\tilde A_I^{(n)} &=& -T_I^{(n)},\\
\tilde B_I^{(n)} &=& -S_I^{(n)}.\nonumber
\end{eqnarray}
This defines the yet undetermined $a$'s and $b$'s, and the equation for
$C_I$ gives one constraint on the $k$ $h_I$ couplings, giving an
additional $(k-1)$-dimensional space of solutions as expected from
the general analysis. Again, this procedure is well defined
and can be extended to any order in perturbation theory,
if the sign of $\hat\gamma$ is right.

\subsection{Special Cases}

The analysis of other special cases is analogous to our analysis
above, and we will present here only the results (details may be found in
\cite{a:Thesis}).

\vskip .2in
 $\bullet$ $k=3$, $\vec{a}=(1,-1,0)$
\vskip .2in

In addition to both types of deformations discussed in the
previous subsection we
can have here also a superpotential
$W={\kappa\over 3!} \Tr(Q_1Q_2Q_3)+{\tilde\kappa\over 3!} \Tr(\tilde
Q_1\tilde Q_2\tilde Q_3)$. In this case we find, both in the general
analysis and in the one-loop analysis, $7$
\textit{exactly} marginal deformations, which is three deformations
more than the general ${\cal N}=2$ case we treated above.

\vskip .2in
 $\bullet$  $k=3$, $\vec{a}=(1,1,1)$
\vskip .2in

In this case we can also have a superpotential of the form
$W=h^I_{ijk}\Tr(Q^I_iQ^{I+1}_jQ^{I+2}_k)$, and we find $3$
\textit{exactly} marginal deformations, one beyond the number of
deformations expected from the generic case.

 \vskip .2in
 $\bullet$  $\vec{a}=(a,a,-2a)$
\vskip .2in

 This case includes theories with ${\cal N}=2$ SUSY
 ($k=2$, $\vec{a}=(1,-1,0)$) and theories with ${\cal
 N}=1$ SUSY. The extra interactions that can be turned on here are
 $h^I_{113},h^I_{223}$ from (\ref{sup_pot}).
We define $p_I\equiv h^I_{113}$ and
$s_I\equiv h^I_{223}$. The extra $\beta$-functions are:
\begin{eqnarray}
   \beta_{s_I}&\propto&\gamma_2^I+\gamma_2^{I+a}+\gamma_3^{I+2a},\nonumber\\
   \beta_{p_I}&\propto&\gamma_1^I+\gamma_1^{I+a}+\gamma_3^{I+2a}.
 \end{eqnarray}
From here one can obtain that in the odd $k$ case all
the anomalous dimensions $\gamma^I_1,\gamma^I_2$ are equal, and in
the even $k$ case we have two independent anomalous dimensions :
$\gamma^{I+2a}_{1,2}=\gamma^{I}_{1,2}$. We also see that all the
$\gamma_3^I$'s are equal and equal to $-(\gamma_1^I+\gamma_1^{I+a})$.

In the even $k$ case we find a new solution, not obtained in
the general case. This solution is parameterized by three couplings,
which can be chosen to be the gauge
coupling (equal for all groups) and two of the $p_I$'s ($p_I=p_{I+2a}$).

\vskip .2in
 $\bullet$  $k=3k'$, $\vec{a}=(a,{k\over
3}-a,-{k\over3})$ and $k=6$, $\vec{a}=(1,1,4)$.
 \vskip .2in

In these cases we can turn on superpotentials of the form
$h^I_{333}\Tr(Q^I_3Q^{I+{2k\over3}}_3Q^{I+{4k\over3}}_3)$. We add
$k'$ couplings but we also add $k'$ new $\beta$-functions,
\begin{eqnarray}\label{spec1}
 \beta_{h^I_{333}}\propto
 \gamma^I_3+\gamma^{I+{k'}}_3+\gamma^{I+2k'}_3=0,
\end{eqnarray}
or $\sum_{I\in {\cal
S}^J_{2k\over3}}\gamma_3^I=0$. Thus we do not expect any new
\textit{exactly} marginal directions here and we indeed don't see
them in the perturbative calculation.

\vskip .2in
 $\bullet$  $SU(N=3)$
\vskip .2in

 As we mentioned, we can have here a larger class of marginal
 deformations (\ref{N3}). Again we discuss the example of the ${\cal
 N}=2$ theory and briefly report the results for other cases.
 In this case one can add additional superpotentials of the form ${\rho_I\over
 3!} \epsilon_{lmn}
 \epsilon^{abc}(Q_I)^l_a(Q_I)^m_b(Q_I)^n_c$ and ${\tilde\rho_I\over
 3!} \epsilon_{lmn} \epsilon^{abc} (\tilde Q_I)^l_a (\tilde
Q_I)^m_b(\tilde Q_I)^n_c$, and
\begin{eqnarray}
  \beta_{\rho_I}=3\rho_I\cdot\gamma_{Q_I},\nonumber\\
  \beta_{\tilde\rho_I}=3\tilde\rho_I\cdot\gamma_{\tilde Q_I}.
\end{eqnarray}
From here and (\ref{betas}) we find for all $I$
$\gamma_{Q_I}=\gamma_{\tilde Q_I}=\gamma_{\Phi_I}=0$.
 In this case adding non zero $h_I$'s doesn't
 change the conditions for the $\gamma$'s
($\gamma_{\Phi_I}$ has to be zero anyway), so we can consider them
 together.
 We have here $6k$ couplings and $3k$
 conditions, leading naively to a $3k$ dimensional manifold
 of fixed points.
 The interactions we add affect the one loop $\gamma_{\Phi_I}$ as in
 (\ref{1_loop_C}), with
 additional terms $2\rho_I^2(\equiv K_I)$ for $Q$ and
 $2\tilde\rho_I^2(\equiv\tilde K_I)$ for $\tilde Q$ :
\begin{eqnarray}
  B_I+A_I&=&-C_I,\nonumber\\
  B_{I+1}+A_I&=&-K_I,\\
  B_{I+1}+A_I&=&-\tilde K_I.\nonumber
\end{eqnarray}
At one loop we find that necessarily
\begin{eqnarray}\label{ccon}
   \sum_I C_I&=&\sum_I K_I,\\
   K_I&=&\tilde K_I,\nonumber
\end{eqnarray}
and the one loop solution is:
\begin{eqnarray}
 |\delta_{I-1}|^2&=&X+4g_I^2+\sum_{J=1}^{I-1}(C_J-K_J),\nonumber\\
 |\alpha_I|^2&=&4g_I^2-X+\sum_{J=1}^{I-1}K_J-\sum_{J=1}^{I}C_J.
\end{eqnarray}
The general solution is parameterized by $k$ gauge couplings, the $X$
 parameter, $k$ $\rho_I$'s and $k$ $h_I$'s subject to the condition
(\ref{ccon}) above,
  giving a total of $3k$ parameters, and a $3k$-dimensional
  manifold of fixed points as expected. These solutions can be
  extended to all loops like we did before.

  The number of
 \textit{exactly} marginal deformations in other cases with
 $SU(N=3)$ gauge group is :

 \begin{itemize}

\item  General case: $3k$ \textit{exactly} marginal deformations.

\item  $k=3$, $\vec{a}=(1,-1,0)$:
11 \textit{exactly} marginal deformations.

\item  $k=3$, $\vec{a}=(1,1,1)$: 33 \textit{exactly} marginal
deformations.

\item  $\vec{a}=(a,a,-2a)$: $5k$ \textit{exactly} marginal deformations.

\item  $\vec{a}=(a,{k\over3}-a,-{k\over3})$, $k=3k'$: $4k$
\textit{exactly} marginal deformations.

\item  $\vec{a}=(1,1,4)$, $k=6$: $36$ \textit{exactly} marginal deformations.

 \end{itemize}

\section{Summary and Discussion}

   First we summarize our results on the \textit{exactly}
marginal deformations :
   \begin{itemize}
   \item ${\cal N}=4$\\
     We found that the only supersymmetric \textit{exactly}
   marginal deformations of ${\cal N}=4$ SYM, other than changing the
gauge coupling, are
   the superpotentials :

    \begin{eqnarray}
      \delta\lambda\epsilon_{ijk}\Tr({\Phi}^i[{\Phi}^j,{\Phi}^k]),\nonumber\\
       \sum_ih\Tr({\Phi}^i\left\{{\Phi}^i,{\Phi}^i\right\}),\\
    h_{123}\Tr({\Phi}^1\left\{{\Phi}^2,{\Phi}^3\right\}),\nonumber
    \end{eqnarray}

  with one equation relating $\lambda$, $h_{123}$, $h$ and the
  gauge coupling.
  These fixed points are IR stable, and the theory
  is not asymptotically free for any choice of the coupling
  constants.

 \item ${\cal N}=2$

   In the ${\cal N}=2$ SYM theories obtained from
   $\mathbb{C}^2/\mathbb{Z}_k$ orbifolds we found :
   \begin{itemize}
   \item General $k$

      We found that there is one \textit{exactly} marginal direction
      (in addition to the $k$ gauge couplings), parameterized by a
      parameter $X$, which can be proven to be \textit{exactly}
marginal at any order of
      perturbation theory. The $X=0$ case has ${\cal N}=2$ SUSY and all
      the $\gamma$'s vanish, however if $X\neq 0$ there may be
      non-zero $\gamma$'s along this direction.
      From the general analysis we expect to have here
      another $k-1$ \textit{exactly} marginal directions, related to
      turning on $\Tr(\Phi_I\Phi_I\Phi_I)$ superpotentials.
      We do not see these \textit{exactly} marginal directions up
      to three loops. This, however, does not necessarily prevent them from
      appearing at higher loops. Whether they appear or not depends on
      the value of a linear combination of the $\gamma_{Q_I}$ and the
      $\gamma_{\Phi_I}$ along the flat direction parameterized by $X$, which
does not include these operators.
      If this linear combination is positive or zero, then these marginal
      directions are ruled out, but if it is negative then these exactly
marginal directions exist.
In any case, the total number of \textit{exactly} marginal
directions is at least $k+1$.

    \item $k=3$

     In this case we have again the gauge couplings and the $X$ deformation,
as above, and we can
     also have three additional
\textit{exactly} marginal directions. The perturbative result
     agrees with the general analysis, and we see all the
     marginal deformations already at one loop. The total
     number of \textit{exactly} marginal directions here is $7$.

   \item $SU(N=3)$

     Here we have yet a larger space of deformations : in addition to
     the gauge couplings and the $X$ deformation,
     we get $2k-1$ additional deformations, for a total of $3k$
     \textit{exactly} marginal deformations for
     general $k$, and $11$ for $k=3$. Again we see all the
     deformations already at one-loop, in agreement with the general analysis.
   \end{itemize}

 \item ${\cal N}=1$

    For the theory coming from a $\mathbb{Z}_k$ orbifold with
    general $(a_1,a_2,a_3)$,
    denoting the largest common divisor of
    $a_i$ with $k$ by $\alpha_i$,
we show that the number
    of \textit{exactly}
    marginal directions is $\sum_i\alpha_i-1$.
    In the case where two of the $a_i$s
    are equal and $k$ is even we get
    additional \textit{exactly} marginal directions.
    In the special case of $SU(N=3)$ we get much larger manifolds of
    fixed points, ranging from dimension $3k$ in the most
    general $(a_1,a_2,a_3)$ $\mathbb{Z}_k$
    theory, to $11k$ in the $k=3$ case.

\end{itemize}

In the large $N$ limit, there is a strong relation between the
orbifold theories and the parent ${\cal N}=4$ theory. Operators in the
${\cal N}=4$ theory which are invariant under the orbifold action are
related to ``untwisted operators'' in the orbifold theories; for
example, the operator $\Tr(F_{\mu \nu}^2)$ in the ${\cal N}=4$ theory
is related to the operator $\sum_I \Tr((F_{\mu \nu}^I)^2)$ in the
orbifold theories. The relation is \cite{a:N_plan1,a:N_plan2} that in
the large $N$ limit, where only planar diagrams contribute to
correlation functions, all correlation functions of these operators
are equal (up to some powers of $k$) in the ``parent'' theory and in
the orbifold theories.

Clearly, this implies that in the large $N$ limit, \textit{exactly} marginal
deformations of the ${\cal N}=4$ theory
which are invariant under the orbifolding should correspond to exactly
marginal deformations of the orbifold theories as well. Our analysis
above shows that this is, in fact, true even at finite $N$. Two of the
\textit{exactly} marginal deformations of the ${\cal N}=4$ theory are always
invariant under the orbifolds we perform; the only one which is not
is $W = h\sum_i \Tr((\Phi^i)^3)$. And indeed, in all our orbifold
theories we find two \textit{exactly} marginal deformations coming from these
deformations. Such \textit{exactly} marginal deformations are necessarily
invariant under the action of the $\mathbb{Z}_k$ orbifold group. For
example, in the ${\cal N}=2$ case, these two deformations are the
equal change in all the gauge couplings, and the deformation
parameterized by $X$. In the special $k=3$ case, the third \textit{exactly}
marginal deformation of ${\cal N}=4$ SYM is also invariant under the
orbifolding, and indeed we find this additional deformation in all the
cases with $k=3$. So, our results are consistent with the expectations
from the orbifold point of view.

The theories we describe here all have string theory duals, given by
type IIB string theory on $AdS_5\times S^5/\mathbb{Z}_k$
\cite{a:Maldac,a:OA}. \textit{Exactly} marginal deformations correspond to
moduli of these string compactifications, in which some massless
scalar fields on $AdS_5$ acquire VEVs. The string theory dual of the
\textit{exactly} marginal deformations of ${\cal N}=4$ SYM will be discussed,
in the large $N$ supergravity approximation, in \cite{a:AKS}.
When the deformation is invariant under $\mathbb{Z}_k$, one
can orbifold also the deformed backgrounds, and this will give the
string theory backgrounds corresponding to the \textit{exactly} marginal
deformations of the orbifold theories. Alternatively, they can be
described by turning on untwisted sector moduli fields in $AdS_5\times
S^5/\mathbb{Z}_k$.

The additional \textit{exactly} marginal deformations that we find in the
   orbifold theories are all related to twisted sector moduli in the
   string theory. Such moduli can only appear on the string theory
   side if the orbifold action has fixed points on the $S^5$, at which
   massless twisted sector fields are located. For example,
in the ${\cal N}=2$ case, the
   orbifold keeps one direction in $\mathbb{C}^3$ fixed,
and thus because the orbifold acts on the $S^5$ factor of the
$AdS_5\times S^5$ as it acts on the angular coordinates of the
$\mathbb{R}^6\sim \mathbb{C}^3$, we have a fixed $S^1$ in the
$S^5$. This enables the
    appearance of massless twisted sector states which can
   correspond to some \textit{exactly} marginal operators on the
field theory side. In this case
   these $k-1$ deformations correspond to $k-1$ blow up
   modes of the $\mathbb{C}^2/\mathbb{Z}_k$ singularity
    \cite{a:SilKach} (more precisely, they correspond to the 2-form
    fields on the vanishing 2-cycles in these singularities).
Another case of \textit{exactly} marginal operators coming from the
twisted sector is
   the $2$ additional operators we get in the ${\cal N}=2$ $\mathbb{Z}_3$ case.

   In the ${\cal N}=1$ case the only fixed point of the full $\mathbb{Z}_k$
action is the
   origin of $\mathbb{C}^3$. However, we still can have massless
   twisted sector states if some of the elements of the orbifold group
have fixed points on the $S^5$. The action of the orbifold is
given by :
   \begin{eqnarray}
  \wp \equiv
   \left(\begin{array}{ccc}
e^{{2\pi i\over k}a_1} &0  &0  \\
0 &e^{{2\pi i\over k}a_2}  &0  \\
0 &0  & e^{{2\pi i\over k}a_3}
\end{array}\right).
\end{eqnarray}
Now, if $\alpha_1$ is the largest common divisor of $k$ and $a_1$,
and if we start with the vector $(1,0,0)$, we will get back to our
starting point after ${k\over \alpha_1}$ applications of ${\bf
\wp}$. So, if $\alpha_1 > 1$, the
  $k\over\alpha_1$'th twisted sector
  has fixed points and could include massless
    states. More generally we find $\alpha_i-1$ twisted sectors which
could have massless states living on a circle on $S^5$.

In the general theory we found $\sum_i \alpha_i-1$ \textit{exactly} marginal
deformations, two of which came from the untwisted sector. Thus, we
see that each twisted sector with fixed points contributes precisely one
\textit{exactly} marginal deformation to the theory, corresponding to turning
on a massless scalar in this twisted sector.
The  $\sum_i\alpha_i-3$  \textit{exactly} marginal deformations
coming from the twisted sectors can be related to the ${\cal N}=2$
blow up modes -- locally near the fixed line where the light twisted sector
states live, the background looks like the ${\cal N}=2$ $AdS_5\times
S^5/\mathbb{Z}_{\alpha_i}$ theory, and these modes will correspond to
2-form fields on vanishing 2-cycles.

So, in the general case we have a nice interpretation for all the
deformations, as coming from untwisted or twisted sectors in the
string theory, with one deformation in each twisted sector with
fixed points. For $SU(N=3)$ we get a much larger space of
deformations. It would be interesting to understand the origin of
this larger space of deformations on the string theory side. Of
course, supergravity is not a good approximation in such a case,
so this would require a full understanding of string theory on
$AdS_5\times S^5/\mathbb{Z}_k$. In some cases we also find
additional \textit{exactly} marginal deformations coming the
twisted sector : the $2$ extra deformations of the ${\cal N}=2$
$\mathbb{Z}_3$ theory and the $(a,a,-2a)$ extra deformations, for
instance. In such cases additional massless twisted sector fields,
beyond the one complex scalar of the general case, should also
correspond to \textit{exactly} marginal deformations. It would be
interesting to understand the form of these \textit{exactly}
marginal deformations on the string theory side.

An interesting point which we have not been able to resolve
involves the marginal operators of the form $\Tr(\Phi_I^3)$ in the
general ${\cal N}=2$ $\mathbb{Z}_k$ case. From the general
analysis, by counting the number of variables and equations, $k-1$
of these operators (which correspond to some twisted sector
fields) are expected to be \textit{exactly} marginal. However, we
saw that these operators do not appear in perturbation theory up
to 3-loop order. Higher loop calculations are required to decide
if these exactly marginal deformations appear in the weak coupling
region or not. It would be interesting also to analyze these
marginal deformations, using string theory, in the strong coupling
regime, and to see if \textit{exactly} marginal deformations
appear there or not.

\vspace{1cm}

\centerline{\bf Acknowledgements}

The research of O.A. was supported in part by
the Israel-U.S. Binational Science Foundation,
by the IRF Centers of
Excellence program, by the European RTN network HPRN-CT-2000-00122,
and by Minerva.

{}

\begin{thebibliography}{99}

\bibitem{a:Maldac} J. Maldacena, ``The large $N$ limit of
superconformal field theories and supergravity",
Adv. Theor. Math. Phys. 2 (1998) 231, hep-th/9711200.
\bibitem{a:OA} O. Aharony, S. S. Gubser, J. Maldacena, H. Ooguri and
Y. Oz, ``Large $N$ field theories, string
theory and gravity", Phys. Rept. 323(2000) 183, hep-th/9905111.
\bibitem{a:quiver} M. R. Douglas and G. Moore,
``D-branes, Quivers, and ALE Instantons", hep-th/9603167.
\bibitem{a:quiver2} M. R. Douglas, B. R. Greene, D.
R. Morrison, ``Orbifold resolution by D-branes", Nucl. Phys. B532
(1998) 163, hep-th/9704151.
\bibitem{a:SilKach} S. Kachru, E. Silverstein, ``4d Conformal Field
Theories and Strings on Orbifolds", Phys. Rev. Lett.
80 (1998) 4855, hep-th/9802183.
\bibitem{a:Leigh} R. G. Leigh and M. Strassler,
``Exactly marginal operators and duality in four dimensional
 $N=1$ supersymmetric gauge theory", Nucl.
Phys. B447 (1995) 95, hep-th/9503121.
\bibitem{a:Leigh1} D. Berenstein, V. Jejjala, R. G. 
Leigh, "Noncommutative moduli spaces, dielectric tori and T duality", Phys.Lett. B493 (2000) 162,
 hep-th/0006168
 \bibitem{a:Leigh2}  D. Berenstein, V. Jejjala, R. G. 
Leigh, "Marginal and relevant deformations of N=4 field theories and 
noncommutative moduli spaces of vacua", Nucl.Phys.B589 (2000) 196, hep-th/0005087 
 \bibitem{a:N_plan1} A. Lawrence, N. Nekrasov, C. Vafa, ``On Conformal
Field Theories in Four Dimensions", Nucl. Phys. B533 (1998) 199,
hep-th/9803015.
\bibitem{a:N_plan2} M. Bershadsky and A. Johansen, ``Large $N$ Limit of
Orbifold Field Theories", Nucl. Phys. B536 (1998) 141,
hep-th/9803249.
\bibitem{a:KleStr} I.~R.~Klebanov and M.~J.~Strassler,
``Supergravity and a confining gauge theory: Duality cascades and
$\chi$SB-resolution of naked singularities,'' JHEP {\bf 0008}
(2000) 052, hep-th/0007191.
\bibitem{a:NSVZ1} V. A. Novikov, M. A. Shifman, A. I. Vainshtein and
V. I. Zakharov, ``The $\beta$-function in supersymmetric gauge
theories. Instantons versus traditional approach", Phys. Lett. B166
(1986) 329.
\bibitem{a:NSVZ2} V. A. Novikov and A. Vainshtein, Nucl. Phys. B277 (1986) 456.
\bibitem{a:NSVZ3} N. Arkani-Hamed and H. Murayama, ``Holomorphy,
rescaling anomalies and exact $\beta$-functions in supersymmetric
gauge theories", JHEP 0006 (2000) 030, hep-th/9707133.
\bibitem{a:Thesis} S. Razamat, ``Marginal deformations of ${\cal N}=4$
SYM and of its supersymmetric orbifold descendants", M.Sc. thesis,
hep-th/0204043.
\bibitem{a:AKS} O.~Aharony, B.~Kol and S.~Yankielowicz, to appear.
\bibitem{a:petrini} L. Girardello, M. Petrini, M. Poratti and
A. Zaffaroni, ``Novel local CFT and exact results on perturbations
of N=4 Super Yang Mills from AdS dynamics", JHEP 9812 (1998) 022,
hep-th/9810126.
\bibitem{a:Reduct}  A. V. Ermushev, D. I. Kazakov and O. V. Tarasov,
``Finite N=1 supersymmetric grand unified theories", Nucl. Phys.
B281(1987) 72.
\end{thebibliography}
\end{document}